\documentclass[12pt,showkeys]{article}
\begin{document}
\renewcommand{\theequation}{\thesection.\arabic{equation}}

\title{$\kappa$-Deformations of $D=3$  Conformal
Versus Deformations of  $D=4$ AdS Symmetries}

\author{J. Lukierski$^{1),2)}$, V.D. Lyakhovsky$^{1), 3)}$
 and M. Mozrzymas$^{1)}$
\\ \\
$^{1)}$Institute for Theoretical Physics, \\
 University of Wroc{\l}aw,
pl. Maxa Borna 9,                             \\
 50--205 Wroc{\l}aw, Poland
\\
$^{2)}$
Departamento de Fisica Teorica\\
Universidad de Valencia, Av. Dr. Moliner 50 \\
46 100 Burjasot (Valencia), Spain
\\  
$^{3)}$Departament of Theoretical Physics, \\
Sankt-Petersburg
University,
         Ulianovskaya 1, Petrodvoretz, \\
198904 Sankt Petersburg, Russia}

\date{}
\maketitle
\thispagestyle{empty}

\begin{abstract}
We describe the classical $o(3,2)$ $r$-matrices as generating the
quantum
 deformations of either D=3 conformal algebra with mass-like deformation
parameters
 or D=4 $AdS$ algebra with dimensionless  deformation parameters.
 We describe the quantization of classical $o(3,2)$ $r$-matrices
   via Drinfeld twist
method which locates the deformation in the coalgebra sector.
 Further we obtain
  the quantum $o(3,2)$  algebra
  in
  a convenient Hopf algebra form
 by considering suitable deformation maps from classical to deformed
  $o(3,2)$ algebra basis. It appears that
   if we pass from $\kappa$-deformed D =3 conformal algebra basis to
   the deformed
      D=4 $AdS$ generators
    basis
    the role
   of dimensionfull parameter  is taken over by the $AdS$ radius $R$.
    We provide also
     the bilinear $o(3,2)$ Casimir  which we express using the
 deformed D=3 conformal  basis.
\end{abstract}

\section{Introduction}
\setcounter{page}{0}

Let us recall  that the nonstandard quantum deformation of
 $sl(2;R)\simeq o(2,1)$ has been
  interpreted in
\cite{ll1}
   as the $ \kappa$-deformation
 of D=1 conformal algebra. Moreover, it appears that if we consider the
 solutions of classical Yang-Baxter equations with support in Borel
 subalgebra  of
  $o(D,2)$ (for $D > 1$), after quantization we obtain the nonstandard
  quantum deformations\footnote{We remind here that standard quantum
  deformation is given by Drinfeld-Jimbo deformation scheme
  \cite{ll2,ll3},
  which
   we shall call $q$-deformation. The $q$-deformation of $o(D,2)$
   implies that the deformation parameter $q$ is dimensionless
   (for D=4 see
   [4,5]).}
   of $D$-dimensional conformal algebra with mass-like
    deformation parameter.
   The $D=2$ case of $o(2,2)$ due to the algebra isomorphism
$o(2,2) \simeq o(2,1)  \oplus o(2,1)$ can be reduced to D=1 case
(see e.g.
 \cite{ll6}). The simplest new case is given by D=3 conformal
algebra
 $o(3,2)$\footnote{The case $D=4$, i.e. nonstandard quantum
 deformations of D=4 conformal
algebra were considered recently by the same authors in
 \cite{ll7}.}.

 The quantum deformation of $o(3,2)$ which can be interpreted as
 \sloppy $\kappa$-de\-for\-ma\-tion of D=3 conformal algebra has been
 first given
  by Herranz
   \cite{ll8}. His result was obtained by checking the Hopf
   algebra relations, with coproduct sector defining
    the classical $o(3,2)$ $r$-matrix by its  lowest order term.
    Our approach is  different: firstly we consider the general
     formula for classical $o(3,2)$   $r$-matrices
    providing $\kappa$-deformations, then we perform
    the quantization using the Drinfeld twist technique
     \cite{ll9}--\cite{ll11}.
     In this  way we obtain the quantum $o(3,2)$  Hopf algebra in
    classical Lie algebra basis and achieve
    better control over the structure of this
    quantum Hopf algebra.
    Further we perform the nonlinear deformation map which provides
     for D=3 conformal algebra
    the deformed algebraic relations,
     but leads to more convenient form of
     coalgebra relations. However one can also interpret the
     quantum algebra $o(3,2)$ as a quantum deformation of D=4 $AdS$
   algebra, with dimensionfull parameter $R$ describing the
   $AdS$ radius. If we reexpress  the
    considered  here $\kappa$-deformations of D=3 conformal
    algebra as describing the deformations of D=4 $AdS$
      symmetry we
    obtain rather surprising
     result that former dimensionfull "$\kappa$-parameters" should
     be considered  as dimensionless.
     Further we shall
     address
    the question
  how to  obtain deformed $o(3,2)$ Casimirs (for
    simplicity we shall consider only bilinear case) which
     we can interpret in D=3
    conformal or D=4 $AdS$ basis.

     The plan of our paper is the following:

     In Sect. 2 we shall classify  in "mathematical"
     Cartan-Weyl basis the solutions of CYBE describing
     the three-parameter family of
     classical $r$-matrices with support in  $B_+ \wedge B_+$ where
     $B_+$ denotes the Borel algebra of $o(3,2)$.
    Using inner automorphisms one can consider the classes of
    mathematically equivalent $r$-matrices. We shall show
    that up to this equivalence the considered three-parameter
    family is the continuous one parameter set of (mathematically)
    nonequivalent classical r-matrices. 
     Subsequently we 
      impose the
    $o(3,2)$ reality conditions for the generators and choose
    respectively conditions for the deformation parameters
    permitting to define the real $o(3,2)$ quantum Hopf algebras.

     We shall calculate the Drinfeld twist and obtain
     explicite formulae providing the coproducts in classical
     $o(3,2)$ basis (see Appendix).
       Further   we  perform the nonlinear
     transformation (deformation map) of $o(3,2)$ basis which
     provides the quantum    deformation of $o(3.2)$  algebra
     in a convenient (Hopf algebraic)
     form\footnote{For the special choice
     of deformation parameter our formulae are
     similar to the ones  given by Herranz
      \cite{ll8}  but not
     identical. We find also some problems with the reality condition for
     deformed D=3 conformal algebra proposed in
     \cite{ll8}.}.
     The new multiplication and coproducts  are
     given by explicite formulae.  Finally, as a byproduct of our
     method the bilinear Casimir in
      deformed quantum $o(3,2)$
algebra  basis     is given.

     In Sect. 3 we shall interpret the result of
     Sect. 2 in physical D=3 conformal and D=4 $AdS$ bases.
      It appears
      that the
        same mathematical
       deformation parameters have different meaning
       in these two frameworks: for D=3 conformal algebra
       they have
       the dimension of the inverse of mass and describe
        $\kappa$-deformation, and for D=4 $AdS$  algebra
       they are  dimensionless.

     The paper is concluded by Sect. 4 with an outlook.
     In this last Section we provide the formula for
     universal quantum $R$-matrix.

\section{Classical $o(3,2)$ $r$-matrices
 and twist quantization of $U(o(3,2))$ in mathematical basis}

\setcounter{equation}{0}

 i) {\bf Classical $o(3,2)$ $r$-matrices in Cartan-Weyl basis}

The simple complex  Lie algebra $B_2 = o(5) \simeq sp(4) = C_2$ has two
simple roots ${\alpha}_1, \alpha_2$ with length  squares one and
two, and symmetrized Cartan matrix ($i=1,2$)
\begin{equation}\label{llm2.1}
  \alpha_{ij} = (\alpha_i , \alpha_j ) = \begin{pmatrix}
  {1 & -1 \cr -1 & 2 }\end{pmatrix} \, .
\end{equation}
The Cartan-Chevaley basis for $B_2 \simeq C_2$
\begin{eqnarray}\label{llm2.2}
 && [h_i , e_{\pm j} ] = \pm
\alpha_{ij}\, e_{\pm j} \, ,
\qquad \qquad [h_i , h_j ] = 0 \, ,
 \cr\cr
  && [ e_{+ i} , e_{- j} ] =
\delta_{ij}\, h_i \, ,
\end{eqnarray}
should be enlarged by the generators $e_{\pm 3}, e_{\pm 4 }$
corresponding to   nonsimple roots $\alpha_3 = \alpha_1 +
\alpha_2$ and $\alpha_4 = \alpha_1 + \alpha_3 $
\begin{equation}\label{llm2.3}
  e_{\pm 3} = \pm [ e_{\pm 1} , e_{\pm 2} ] \, ,
  \qquad
  e_{\pm 4} = \pm [ e_{\pm 1} , e_{\pm 3} ] \, ,
\end{equation}
which together with the Chevalley basis (\ref{llm2.2})
describe, due to Serre relations, the whole
10-generator Lie algebra $B_2 \simeq C_2$.
 It can be checked that this algebra is invariant under
 the following  complex rescaling of generators:
\begin{equation}\label{llm22.4}
  h'_l = h_i \, , \qquad e'_{\pm 1} = e_{\pm 1} \, , \qquad
  e'_{\pm l} = \lambda^{\pm 1} e_{\pm l} \, , \qquad l=2,3,4 \, ,
\end{equation}
which  for the real form $o(3,2)$ ($\lambda$ real)
 will be interpreted physically in Sect. 3.

In order to obtain the real form $o(3,2)$ of $o(5,C)$ we shall use
the Hermitian conjugation map\footnote{Such a map is defined e.g.
for defining matrix representation of
 $o(5,C)$ as well as for the differential realization describing
 the infinitesimal D=5 rotations.}  which leaves invariant the Borel
subalgebra $B^+ = (h_i, e_A)$  ($A=1, \ldots 4)$.
 The $o(3,2)$ generators satisfy the following reality conditions
\begin{equation}\label{llm2.5}
  h^+_i = - h \, , \qquad \qquad e^+_{\pm i} = \lambda_i \, e_{\pm
  i} \, ,
\end{equation}
where $\lambda_i = \pm 1$
  \cite{ll3,ll1}
  that implies $e^+_{\pm 3}= -
\lambda_1 \lambda_2 e_{\pm 3}$ and $e^+_{\pm 4}= - \lambda_2
e_{\pm 4}$.

Smooth triangular quantum deformations of $o(5,C)$ are described
infinitesimally  by classical $r$-matrices, satisfying classical
Yang-Baxter equation.
 We have obtained the following set of three-parameter
  classical $r$-matrices with support in $B^+
\wedge B^+$.
\begin{equation}\label{llm2.6}
  r(\alpha,\xi, \rho) = \alpha [( 2h_1 + h_2 ) \wedge e_4 + 2e_1 \wedge
  e_3] +
  \xi \, h_2 \wedge e_2 + \rho \, e_2 \wedge e_4 \, .
\end{equation}
The invariance of (\ref{llm2.6}) under rescaling (\ref{llm22.4})
implies that
\begin{equation}\label{llm22.7}
  \alpha' = \lambda^{-1} \, \alpha \, , \qquad \xi'= \lambda^{-1}
  \, \xi \, , \qquad \rho'= \lambda^{-2} \, \rho \, .
\end{equation}
We shall show in Sect.3 that the scaling properties
(\ref{llm22.7})
 determine D=3 mass dimensions and
 imply the interpretation of the deformation
(\ref{llm2.6}) as representing D=3 $\kappa$-deformations.
Due to
the most general  two-parameter scaling
   automorphisms
    of $o(5,C)$ algebra\footnote{The number of parameters for
    such scale automorphisms is equal to the rank of the Lie algebra}
\begin{equation}
\begin{array}{cccc}
e_{1}\Rightarrow \alpha ^{\frac{1}{2}}\xi ^{-\frac{1}{2}}e_{1}, &
e_{-1}\Rightarrow \alpha ^{-\frac{1}{2}}\xi ^{\frac{1}{2}}e_{-1},
& e_{2}\Rightarrow \xi e_{2}, &
e_{-2} \Rightarrow \xi ^{-1}  e_{-2} \, ,%
\\
e_{3}\Rightarrow \alpha ^{\frac{1}{2}}\xi ^{\frac{1}{2}}e_{3}, &
e_{-3}\Rightarrow \alpha ^{-\frac{1}{2}}\xi ^{-\frac{1}{2}}e_{-3},
& e_{4}\Rightarrow \alpha e_{4}, & e_{-4}\Rightarrow \alpha
^{-1}e_{-4}\, .
\end{array}
\label{llm2.14}
\end{equation}%
  two out of
three parameters $\alpha , \xi, \rho$ can be fixed in a particular
way.

If we choose in (\ref{llm2.5}) $\lambda_i = -1$ all 10 generators
 describing $o(3,2)$ are anti-Hermitian. One gets
 in formula (\ref{llm2.6})
  the Hermitian classical $o(3,2)$ $r$-matrices
    if the parameters $\alpha, \xi, \rho $ are  real;
    the purely  imaginary choice of of $\alpha, \xi, \rho$ implies
    that the classical $r$-matrices are anti-Hermitian.
    Further we shall employ the second choice, which will provide
    after twist quantization the
    real quantum $o(3,2)$ Hopf algebras.

Below we shall assume that $\rho =0$\footnote{The
 third term in (\ref{llm2.6}), proportional to $\rho$, introduces
 Reshetikhin twisting factor on Abelian subalgebra
 ($[e_2, e_3] =0$). Such terms in classical $r$-matrix
 generate "soft deformations".}, i.e. the
 substitution (2.8) can transform the parameters $\alpha = 1$,
 $\xi=1$ in (2.6) and attach to them arbitrary nonzero values.
The example of $\kappa$-Poincar\'{e} algebra
  \cite{ll2}--\cite{ll14}
 shows that
 such deformations being equivalent
 mathematically are
 not equivalent physically, so further
 we shall keep $\alpha$ and $\xi$ arbitrary.

 Finally we observe that the
  matrix (\ref{llm2.6}) for $\xi=\rho=0$ was presented in
   \cite{ll1},
and the choice corresponding to $\alpha= \xi= \frac{1}{2}$,
$\rho=0$  was obtained by Herranz
 \cite{ll8}.

ii) {\bf Twisting Elements and their Parametrizations}

Twist deformations of enveloping algebra $U(g)$
 are defined by the
twisting elements ${\Phi }=\sum f_{\left( 1\right) }\otimes
f_{\left(
2\right) }\in U(g)\otimes U(g)$ that satisfy the twist equations
 \cite{ll9}%
:
\begin{equation}\label{llm2.9}
\left( {\Phi }\right) _{12}\left( \Delta \otimes {\rm id}\right) {\ \Phi }%
=\left( {\ \Phi }\right) _{23}\left( {\rm id}\otimes \Delta \right) {\ \Phi }%
,\qquad \left( \epsilon \otimes {\rm id}\right) {\ \Phi }=\left( {\rm id}%
\otimes \epsilon \right) {\ \Phi }=1.  \label{twist-eq}
\end{equation}%
The quantized algebra $U_{{\ \Phi }}(g)$ has the same commutation
relations as the classical enveloping algebra   $U(g)$, the
deformation is present only in the new
coproducts that can be obtained in the form
\begin{equation}\label{xxx}
 \Delta _{{\ \Phi }}={\ \Phi }
\Delta {\ \Phi }^{-1}\, ,
\end{equation}
and in the antipode formulae.
Recently the effective methods to find explicit solutions to
(2.9) were elaborated
 \cite{ll10,ll11}
 for classical simple Lie
algebras. In particular the deformed carrier space algorithm was
discovered
  \cite{ll15}
 and chains of twists for orthogonal algebras
were constructed in
 \cite{ll16}, where the technique of
finding the factorized solution with carrier  in
$B^{+}(o(M))$ is given.

Applying it to $U(o(3,2))$ one can check that the two-parameter
twisting element for arbitrary parameters $\alpha$ and $\xi$ in
(2.6) is described by the product of canonical twist element
(corresponding to $\xi = \rho = 0$) and deformed Jordanian twist
(corresponding to $\alpha = \rho = 0$). One gets the formula:
\begin{equation}\label{llm2.15}
{\cal F}(\alpha, \xi)=\exp \frac{1}{2} \left( h_{2}
\otimes {\bf \sigma
}_{2}{(\alpha, \xi)}\right) \exp \left( \alpha e_{1}
\otimes
e_{3}\,
e^{-\frac{1}{2}{ \sigma }_{4}(\alpha)}\right) \exp
\frac{1}{2}\left( h_{4}\otimes {\bf \sigma }_{4}{(\alpha)}\right)
, \label{param-tw}
\end{equation}%
Notice that here the formal power series
 ${ \sigma_2, \sigma_4 }$ are now
functions
of $\alpha $ and $\xi $%
\begin{eqnarray}\label{llm2.16}
{\bf \sigma }_{4}{(\alpha)}
&=&
 \ln \left( 1 + \alpha e_{4}\right) ,
\label{sigmas}
\\
{\bf  \sigma }_{2} (\alpha, \xi )  &=&\ln \left( 1- \xi
e_{2}+\frac{1}{2}
\alpha \, \xi \, e_{3}^{2}\, e^{-{\bf \sigma }_{4}
(\alpha)}\right) .  \nonumber
\end{eqnarray}%

iii) {\bf Deformed twisted $o(3,2)$ coproducts}

 The universal enveloping algebra $U(o(3,2))$ is not modified
by twisting procedure and   the whole deformation
is located in the coalgebra sector. The costructure and
correspondingly the
tensor product rules are defined by the new coproducts according
to the formula (\ref{xxx}). We present the explicit form of these
coproducts in the Appendix.

It can be shown that for imaginary parameters $\alpha$ and $\xi$
the coproduct map is real, i.e. $\Delta(a^+)=(\Delta(a))^+$, where
$a\in U(o(3,2))$ and $(a \otimes b)^+= a^+ \otimes b^+$.

iv) {\bf The nonlinear deformation map and quantum $o(3,2)$ algebra in
nonclassical basis}

We see from the formulae in the Appendix that the generators
$(h_i , e_A, ) \subset B^+$ have considerably simpler coproducts
than the generators $e_{-A} \in B^-$.
 According to the duality defined in
 $U_{\cal F}$ by the twisting element ${\cal F}$ (see the deformed
 carrier space approach in
  \cite{ll16}) the appropriate generators for the
 carrier subalgebra are just the tensor multipliers in
 $\ln ({\cal F})$. In our case these are $h_2,h_4, \sigma_2, \sigma_2,
 e_1$ and $e_3 e^{-\sigma_4}$. The remaining four elements must be
 correlated with the obtained new positive root generators $(E_1,
 \dots E_4 )$ to conserve the relations $ [E_{p}, E_{-p} ]=H_p $.
 Finally we get the following deformed basis:

\begin{eqnarray}\label{llm2.17}
&& H_2= h_2 \, , \qquad \qquad \qquad\qquad
H_4 = h_4 \, ,
\cr
&&
E_4 =  \alpha^{-1}
 \ln (1 +
\alpha \, e_4 ) \, ,
 \cr
  && E_2 = - \xi^{-1}
 \ln \left[1 -
\xi \, e_2 +
 \frac{1}{2} \alpha \xi  \,
  e^2_3 \left(1 +
\alpha \, e_4
\right)^{-1}\right] \, ,
\cr
&&
 E_3 = e_3 \cdot
\left( 1 +
 \alpha \, e_4
 \right)^{-1} \, ,
 \cr
  &&
  E_1 = e_1 \, ,
\qquad \qquad \qquad\qquad
E_{-1} = e_{-1}  + \alpha \, \xi^{-1} \, e_1\, ,
\cr
 &&
  E_{-2} = e_{-2}
+ \frac{1}{4} \xi \, h^2_2 - \frac{1}{2} \alpha \, e^2_1\, ,
\cr &&
E_{-3} = e_{-3} -
\frac{1}{2} \alpha \,  (h_2 + h_4) e_1 \, ,
\cr &&
E_{-4} = e_{-4}
+ \alpha \xi^{-1}\, e_{-2} +
 \frac{1}{4} \alpha\, h^2_2
  - \frac{1}{4}\alpha \,
h^2_{4} \, .
\end{eqnarray}

It is easy to check that for imaginary parameters $\alpha$ and
$\xi$ the deformed generators (2.14) remain anti-Hermitian.

The set of generators (\ref{llm2.17}) is chosen in such a way that
it provides in the  particular limit ($\alpha \to 0$, $\xi \to 0$;
$\alpha \xi^{-1} \to 0$) the classical generators of $o(3,2 )$
and leads to simplified structure of coproducts\footnote{If we
wish to define te generators having finite limit in  the limit
$\alpha \to 0$, $\xi \to 0$ for any order then one has to redefine
the generators $E_{-1}$ and $E_{-4}$ in (\ref{llm2.17}) }.
Under   rescaling (\ref{llm22.4}) and (\ref{llm22.7}) the deformed
generators (\ref{llm2.17}) transform in the same way as the
corresponding undeformed  classical generators
\begin{equation}\label{llm22.18}
  H'_{i} = H_i \, , \qquad E'_{\pm1} = E_{\pm1} \, , \qquad E'_{\pm l}
  = \lambda^{\pm 1} \, E_{\pm l} \, , \qquad (l =2,3,4)\, ,
\end{equation}
i.e. they have the same mass dimensions as the undeformed generators.

 The coproducts of the
 generators (\ref{llm2.17}) are the following:

\begin{equation}
\begin{array}{lcl}
\Delta _{F}\left( H_{4}\right)  & = & \left(
\begin{array}{c}
H_{4}\otimes e^{-\alpha { E }_{4}}+1\otimes H_{4}-
\\
-\alpha \, E_1 \otimes E_3 \, e^{-\frac{1}{2}
\left( \alpha \,{ E }_{4}-\xi{ E }%
_{2}\right) }-\frac{1}{2}\alpha \xi
 \,H_{2}\otimes E_3^{2}e^{\xi{ E }_{2}}%
\end{array}%
\right) ,
 \\
\Delta _{F}\left( H_{2}\right)  & = & H_{2}\otimes e^{\xi{ E }%
_{2}}+1\otimes H_{2},
\\
\Delta _{F}\left( { E }_{4}\right)  & = & { E }%
_{4}\otimes 1+1\otimes { E }_{4},
\qquad\qquad\qquad
\Delta _{F}\left( {E }_{2}\right)   =  { E } _{2}\otimes 1+1\otimes { E }_{2},
\\
\Delta _{F}\left( E_3\right)
 & = & E_3\otimes e^{\frac{1}{2}\left(- \xi {E }%
_{2}- \alpha { E }_{4}\right) }+1\otimes E_3,
\\
\Delta _{F}\left( E_1\right)
 & = &
 E_1\otimes e^{-\frac{1}{2}\left(  \alpha { E
}_{4}- \xi { E }_{2}\right) }+1\otimes E_1- \frac{1}{2} \xi
H_{2}\otimes E_{3} e^{\xi { E }_{2}}\, ,
\\
\Delta _{F}\left( E_{-1}\right)
  & = & E_{-1}\otimes e^{\frac{1}{2}\left(- \xi { %
E }_{2}- \alpha { E }_{4}\right) }+1\otimes E_{-1},
\\
\Delta _{F}\left( E_{-2}\right)  & =
& E_{-2}\otimes e^{ \xi { E }%
_{2}}+1\otimes E_{-2}, \\
\Delta _{F}\left( E_{-3}\right)  & = & \left(
\begin{array}{c}
E_{-3}\otimes e^{-\frac{1}{2}\left( \alpha { E }_{4}
- \xi { E }%
_{2}\right) }+1\otimes E_{-3}- \\
-  \frac{1}{2} \xi H_{2}\otimes E_{-1}e^{-\alpha E_{4}}
+  \alpha \left(\frac{1}{4}\xi H_2 - E_{-2}\right)
\otimes E_3 e^{ \xi { E }_{2}}%
\end{array}%
\right) , \\
\Delta _{F}\left( E_{-4}\right)  & = & \left(
\begin{array}{c}
E_{-4}\otimes e^{- \alpha { E }_{4}}
+1\otimes E_{-4}+ \alpha E_1\otimes E_{-1}e^{-\frac{%
1}{2}\left(  \alpha { E }_{4} -\xi { E }_{2}\right) } \\
- \alpha E_{-3}
 \otimes E_3 e^{-\frac{1}{2}\left(  \alpha {E
}_{4}- \xi { E }_{2}\right) }+\frac{1}{2}  \alpha \xi
H_{2}\otimes E_{-1}E_3e^{ \xi { E }_{2}} \\
+\left( - \xi^{-1} \alpha E_{-2}+\frac{1}{4}  \alpha H_{2}\right)
\otimes \left( e^{- \alpha { E }%
_{4}}-e^{\xi { E }_{2}} -\frac{1}{2} \alpha \xi E_3^{2}e^{ \xi {
E }_{2}}\right)
\end{array}%
\right) .%
\end{array}%
\end{equation}

 The set of generators (\ref{llm2.17}) satisfy the following set of
 deformed commutators (we write down only nontrivial commutators):

\begin{eqnarray*}
\label{lulymo1}
\begin{array}{rlrl}
    [ H_2 , E_1 ] = & - E_1 \, ,
&
[ H_2 , E_{-1} ]  =& - 2\frac{\alpha}{\xi} E_1 + E_{-1}\, ,
\cr
 [ H_2 , E_2 ] =&- \frac{2}{\xi}\left(1- e^{\xi E_2} \right) \, ,
& [ H_2 , E_{-2} ] =&  - 2E_{-2}- \frac{1}{2} \xi H_{2}^2 \, ,
\cr [ H_2 , E_3 ] =& E_3 \, ,
&[ H_2 , E_{-3} ] =& -  E_{-3} \, ,
\cr [ H_2 , E_{-4} ] =&
    - 2\frac{\alpha}{\xi} E_{-2}  + \frac{1}{2}\alpha H^2_2
    - \frac{\alpha^2}{\xi}\, E_1^2 \, , &
\cr [ H_4 , E_1 ] =& E_1 \, , & [ H_4 , E_{-1} ] = &
2\frac{\alpha}{\xi} E_1 - E_{-1} \,, \cr [ H_4 , E_2 ] =&- \alpha
E_3^2 e^{\xi E_2} \, ,
 &[ H_4 , E_{-2} ]=& -\alpha E_1^2 \, ,
\cr
 [ H_4 , E_3 ] =& 2 E_3 \left( e^{-\alpha E_4} -
    \frac{1}{2}\right)\, ,
\cr
 [ H_4 , E_{-3} ] =& - E_{-3} - \alpha (H_2 +H_4) E_1 \, ,&&
\cr
 [H_4 , E_4 ]   = & \frac{2}{\alpha}
    \left(1-  e^{-\alpha E_4} \right)\, ,&&
\cr
 [ H_4 , E_{-4} ] =& -\frac{\alpha}{2} H_4^2
    + \frac{\alpha^2}{\xi}\, E_1^2
    + 2(\frac{\alpha}{\xi}E_{-2} - E_{-4}) \, ,&&
\cr
 [ E_1 , E_2 ] =&  E_3 \, e^{\xi E_2} \, ,
\qquad
& [ E_1 , E_{-1} ] =& \frac{1}{2} \left( H_4 - H_2 \right) \, ,
\cr [ E_1 , E_3 ] =&\frac{1}{\alpha}
    \left( 1 - e^{-\alpha E_4} \right)\, ,
    \qquad
& [ E_1 , E_{-2} ] =&  \frac{\xi}{2}\left(  H_2 +\frac{1}{2} \right)E_1  \, ,
\cr [ E_1 , E_{-3} ] =& - (E_{-2} -\frac{\xi}{4} H^2_2
    + \frac{1}{2}\, \alpha  E_1^2 )\, ,
    \qquad
    &[ E_1 , E_{-4} ] =& - E_{-3} \, ,
\end{array}
\end{eqnarray*}

\begin{equation}
\begin{array}{rrl}
\cr [ E_2 , E_{-1} ]& =& - \frac{\alpha}{\xi} E_3 \, ,
\qquad \qquad \qquad\qquad\qquad \qquad
 [ E_2 , E_{-2} ] = - H_2 \, ,
\cr
 [ E_2 , E_{-3} ]& =& - ( \frac{\alpha}{2}E_3 - E_{-1})e^{\xi E_2}
+\frac{\alpha}{\xi} E_1 \, ,
\cr [ E_2 , E_{-4} ]& =&  \frac{\alpha}{\xi} H_2
    -  ( \alpha E_{-1} - \frac{1}{4} \alpha^2 E_3)\, E_3 \,
    e^{- \alpha E_4} - \frac{\alpha}{2 \xi} (e^{-\alpha E_4}
    - e^{\xi E_2}) \, ,
\cr [ E_3 , E_{-1} ]& =& \frac{1}{\xi}
    \left( e^{-\xi E_2 - \alpha E_4}-1 \right)
    + \frac{\alpha}{2} E^2_3  \, ,
\cr [ E_3 , E_{-2} ]& =& E_1
    +\xi \frac{1}{2}(\frac{1}{2}-H_2 )\, E_3 \, ,
\qquad\qquad
[ E_3 , E_{-3} ] =  \frac{1}{2} \left( H_2 + H_4 \right)\, ,
\cr [ E_3 , E_{-4} ]& =& E_{-1}e^{-\alpha E_4} -\frac{\alpha}{2}
    (H_2 + H_4)\, E_3 \, ,
\qquad\qquad [ E_4 , E_{-1} ] =  - E_3 \, ,
\cr [ E_4 , E_{-3} ]& =& E_1 \, ,
\qquad\quad\qquad\qquad \qquad\qquad\qquad \qquad [ E_4 , E_{-4} ] = H_4 \, ,
\cr [ E_{-1} ,E_{-2} ]& =& - E_{-3} - \frac{1}{2} \alpha E_1 +
    \frac{1}{2} \xi (\frac{1}{2} - H_2 )\, E_{-1} \, ,
\qquad
 [ E_{-1} , E_{-3} ] =  - E_{-4} \, ,
\cr [ E_{-1} , E_{-4} ]& =& -2\frac{\alpha}{\xi} E_{-3}
    - \frac{\alpha}{2}(H_2 + H_4)\, E_{-1} \, ,
\cr
 [ E_{-2} , E_{-3} ] &=& - \xi \frac{1}{2}(H_2 + \frac{1}{2})\, E_{-3} \, ,
\cr [ E_{-2} , E_{-4} ]& =&  \alpha E_{-3} E_1 + \frac{\alpha}{2}
    (-E_{-2} + \frac{\xi}{4}H^2_2 -\frac{1}{2}\alpha E^2_1) \, ,
\cr [ E_{-3} , E_{-4} ]& =&  \alpha H_2 \, E_{-3} - \alpha E_{-4} E_1 \, ,
\end{array}
\end{equation}

It can be shown that the rhs of relations (2.16) are
anti-Hermitian, in consistency with the anti-Hermiticity of
generators (2.13).

In order to describe completely the Hopf algebra structure of
 deformed $o(3,2)$ we
present the formulae for antipodes:
\begin{equation}
\begin{array}{lcl}
S_{F}\left( H_{4}\right)  & = & \left( -H_{4}-\alpha \,E_{1}E_{3}\,+\frac{1}{%
2}\alpha \xi \,H_{2}E_{3}^{2}\right) e^{\alpha {E}_{4}}, \\
S_{F}\left( H_{2}\right)  & = & -H_{2}e^{-\xi {E}_{2}}, \\
S_{F}\left( {E}_{4}\right)  & = & -{E}_{4}, \\
S_{F}\left( {E}_{2}\right)  & = & -{E}_{2}, \\
S_{F}\left( E_{3}\right)  & = & - E_{3}e^{\frac{1}{2}\left( \alpha {E}%
_{4}+ \xi {E}_{2}\right) }, \\
S_{F}\left( E_{-1}\right)  & = & -E_{-1}e^{\frac{1}{2}\left( \alpha {E%
}_{4} +\xi {E}_{2}\right) }, \\
S_{F}\left( E_{1}\right)  & = & \left( -E_{1}+\frac{1}{2}\xi
H_{2}E_{3}\right) e^{\frac{1}{2}\left( \alpha {E}_{4} - \xi {E}%
_{2}\right) }, \\
S_{F}\left( E_{-2}\right)  & = & -E_{-2}e^{-\xi {E}_{2}}, \\
S_{F}\left( E_{-3}\right)  & = & \left(
\begin{array}{c}
 -E_{-3}-\alpha \left( E_{-2}
-\frac{1}{4}\xi H_2 \right)E_{3} \\
+\frac{1}{2%
}\xi H_{2}\left( E_{-1}-\alpha E_{3}\right) e^{\left( \alpha {E}%
_{4}+\xi {E}_{2}\right) }
\end{array}
\right) e^{\frac{1}{2}\left( \alpha {E%
}_{4} - \xi {E}_{2}\right) }, \\
S_{F}\left( E_{-4}\right)  & = & \left(
\begin{array}{c}
-E_{-4}+\alpha E_{1}E_{-1}- \alpha E_{-3}E_{3}
-\frac{1}{2}\alpha \xi H_{2}E_{3}E_{-1}
\\
+\left( \xi ^{-1}\alpha E_{-2}-\frac{1}{4}\alpha H_{2}\right) \left(
1-e^{\left(- \xi {E}_{2}-\alpha {E}_{4}\right) }-\frac{1}{2}%
\alpha \xi E_{3}^{2}\right)
\end{array}
\right) e^{\alpha {E}_{4}}.
\end{array}
\end{equation}
The counits remain classical.

v) {\bf Deformed Bilinear Casimir}

The classical Casimir operator of $o(5)$
has the form
($r,s=1,2; \, \alpha_{rs}$ is the symmetric Cartan matrix).
\begin{equation}\label{b}
  C_2 = \alpha_{rs} \, h_r \, h_s + \frac{1}{2} (e_A e_{-A} +
  e_{-A} e_{A} ) = h^2_2 + h^2_4 - 2h_4
  -h_2 + e_A e_{-A}  \, .
\end{equation}
In order to apply the formula (2.18) to o(3,2) we should impose
the proper reality conditions (see (2.5))
If we introduce the inverse deformation map
 (i.e. inverse the  relations (\ref{llm2.17}))
 the Casimir operator (\ref{b})
can be expressed in terms of deformed $o(3,2)$
 generators (\ref{llm2.17}) as follows.
\begin{eqnarray}\label{lulymo2.24}
C &  = & 2H^2_1 + H^2_2 + 2H_1\,H_2 + 4 H_1 + 3H_1 +
2 \Big\{ E_1\, E_{-1} - \alpha \xi^{-1} E^2_1
\cr
&& + \
\Big[ \frac{1}{\xi} (e^{-\xi E_2} - 1)
- \frac{1}{2} \alpha \, E^2_3 \,e^{-2\alpha E_4}
[2-e^{\alpha E_4} ]^{-1} \Big]
\left( E_{-2} - \frac{1}{4} \xi \,H^2_2 +
\frac{1}{2} \alpha E^2_1 \right)
\cr
&&+ \ E_3 e^{\alpha E_4} \left( E_{-3} + \alpha H_3 E_1\right) +
\frac{1}{\alpha} \left( e^{\alpha E_4} - 1\right)
\cr
&& \cdot \
\left( E_{-4}
  - \alpha \, \xi^{-1}
\left( E_{-2} + \frac{1}{2}  \alpha \,
E^{2}_1 \right) + \frac{1}{4} \alpha \, H^2_4 \right) \Big\}\, .
\end{eqnarray}

\section{Physical Bases:  D=3 Conformal and
 D=4 $AdS$}
\setcounter{equation}{0}

  i) {\bf D=3 conformal basis}

  Let us introduce purely imaginary $o(3,2)$ generators $M_{\mu\nu} = -
  M^{+}_{\mu\nu} $ ($\mu,\nu = 0,1,2,3,4$) by the relation
  \begin{equation}\label{llm22.9}
  [M_{\mu\nu} , M_{\rho \tau} ] = g_{\mu\tau} \,
  M_{\nu\rho} - g_{\nu\tau} M_{\mu\rho} + g_{\nu\rho}
  M_{\mu\tau} - g_{\mu\rho} M_{\nu\tau} \, ,
  \end{equation}
  where $o(3,2)$ metric has the form          $g_{\mu\nu} = diag
  (- + + - +)$.

  The D=3 Lorentz generators are
  \begin{equation}\label{llm22.10a}
  L_1 = M_{10} \, , \qquad
  L_2 = M_{02} \, , \qquad J=M_{21} \, .
\end{equation}
The threemomenta $P_r , K_r$ ($r=0,1,2$) and dilatation
 generator $D$
are given by the formulae
\begin{equation}\label{llm22.10b}
P_r = \frac{1}{\sqrt{2}} (M_{3r} - M_{4r} ) \, , \qquad K_r =
\frac{1}{\sqrt{2}} (M_{3r} + M_{4r} ) \, , \qquad D= M_{34} \, .
\end{equation}

The relation between CW basis ($h_i , e_{\pm A}$) for $o(3,2)$
and the generators
 $M_{\mu\nu}$  is the following
 \begin{equation}\label{llm2.11}
 \begin{array}{l}
 M_{10} = h_1 \, , \qquad M_{34} = h_3 \, ,
\qquad
  M_{02} = \frac{1}{\sqrt{2}} (e_1 +e_{-1} ) \, ,
 \cr
 M_{32} = \frac{1}{\sqrt{2}} (e_3 +e_{-3} )\, ,
 \quad
  M_{12} = -\frac{1}{\sqrt{2}} (e_1 -e_{-1} ) \, ,
  \quad
   M_{24} = \frac{1}{\sqrt{2}} (e_3 - e_{-3} )\, ,
   \cr
    M_{04} = -\frac{1}{{2}} (e_2 + e_{-2}  + e_4 + e_{-4} )\, ,
\qquad
 M_{14} = -\frac{1}{{2}} (e_2 -  e_{-2}  - e_4 + e_{-4} )\, ,
 \cr
 M_{30} = -\frac{1}{{2}} (e_2 -  e_{-2}  + e_4 - e_{-4} )\, ,
\cr
 M_{31} = -\frac{1}{{2}} (e_2 +  e_{-2}  - e_4 - e_{-4} )\, .
 \end{array}
 \end{equation}

The algebra (\ref{llm22.9}) takes the  form of
D=3 conformal algebra
 ($r,s,u,v=0, 1,2;$ $ g_{rs} = diag (-1,1,1)$):

\begin{equation}
 \begin{array}{l}
[M_{rs}, M_{u v}]  =
 g_{rv} M_{su} -
 g_{ru} M_{sv} + g_{su} M_{rv} - g_{sv}M_{ru}\, ,
\cr
 [M_{rs}, P_{u} ]  =
g_{su}P_r-g_{ru}P_s\, ,
 \cr
[M_{rs}, K_{u} ]  =  g_{ru}K_s-g_{su}K_r\, ,
\cr
[M_{rs} , D] = 0 \, ,
\cr
[D, P_r ]  = - P_r\, , \qquad  \qquad [P_r, P_s ]  = 0 \, ,
\cr
[D, K_r ] =  K_r\, ,  \qquad \qquad [K_r, K_s ] = 0 \, ,
\cr
[K_r, P_s ] =   g_{rs} D + M_{rs} \, .
 \end{array}
\end{equation}

The general $r$-matrix (\ref{llm2.5}) is given in D=3 conformal basis
(\ref{llm22.10a}--\ref{llm22.10b}) by
the formula
\begin{equation}\label{llm22.14}
  r(\alpha, \xi, \rho) = { \alpha}
   [(D+L_1 ) \wedge P_- +
 {\sqrt{2}} (L_2 -J )\wedge P_2]
   - \xi(D- L_1)\wedge P_+ + \rho P_+ \wedge P_-
\end{equation}
where $P_{\pm} = \frac{1}{\sqrt{2}}(P_1 \pm P_0)$.
One can show that the invariance of $r$-matrix (\ref{llm22.14}) under the scale
transformation
\begin{equation}
\label{lulymo3.7}
  P'_r = \lambda^{-1} \, P_r \, , \qquad
  K'_r = \lambda \, K_r \, , \qquad
  M'_{rs} = M_{rs} \, , \qquad D' = D\, ,
\end{equation}
implies that the deformation parameters $\alpha ,\xi, \rho$
  are dimensionfull in accordance with (2.7).
 We see therefore that the
  classical $r$-matrix (\ref{llm2.6}) describes the generalized
    $\kappa$-deformation of D=3 conformal algebra. Introducing the fundamental
    mass parameter  $\kappa$ one can write
\begin{equation}\label{lulymo3.8}
  \alpha = \frac{1}{2\kappa} \, , \qquad
  \xi = \frac{\gamma}{2\kappa} \, , \qquad
  \rho = \frac{\delta}{\kappa^2} \, ,
\end{equation}
The arbitrary choices of the dimensionless parameters $\gamma, \delta$
correspond to generalized
 $\kappa$-deformation of D=3 conformal algebra.


If we put  $ \gamma = 1$, i.e.
$\alpha = \xi = \frac{1}{2\kappa}$ and
 $\rho = 0$ one gets in our D=3 conformal basis
  (3.2--4) the $r$-matrix proposed  by Herranz
   \cite{ll8}
\begin{equation}\label{llm22.15}
  r^H (\frac{1}{2},\frac{1}{2}, 0) = L_1 \wedge P_1 - D \wedge
  P_0 + (L_2 - J ) \wedge P_2 \, .
\end{equation}

In order to rewrite the results
 of Sect. 2 in the framework of D=3 conformal
 algebra we should invert the relations (\ref{llm2.11}). One gets
  (we put $K_{\pm} = \frac{1}{\sqrt{2}} (K_1 \pm K_0)$)

\begin{equation}
 \label{lulymo3.11}
 \begin{array}{ll}
h_1 = H_{12} = L_1\, , \qquad
& h_3 = D \, ,
\cr
e_1 =\frac{1}{\sqrt{2}} (L_2 + J) \, ,
\qquad
 & e_{-1}= \frac{1}{\sqrt{2}} (L_2 -J)\, ,
 \cr
e_2 =
\, P_{+}
 \, ,
& e_{-2} =
 \, K_{-}
 \, ,
\cr
e_3 = P_2 & e_{-3} = K_2 \, ,
\cr
e_4  =
\,  P_{-}
\, ,
& e_{-4}  =
 \, K_{+}
 \, ,
 \end{array}
 \end{equation}

We assume further that the deformed
 nonclassical D=3 conformal generators $\widehat{J},
\widehat{L}_r, \widehat{D},\widehat{P}_l, \widehat{K}_l$ are related with the
deformed generators \sloppy $H_2, H_4, E_{\pm A}$ \sloppy ($\hbox{A=1\ldots 4}$)
 by the same relations (\ref{llm2.11}) and (\ref{lulymo3.11}).
The deformation map (\ref{llm2.17}))
 in terms of D=3 conformal generators looks as follows

\begin{eqnarray}\label{lulymo3.11a}
  \widehat{J}&= &
  \frac{1}{2} [ (2 -\alpha \xi^{-1} ) J - \alpha\xi^{-1}
  L_2  ] \, , \qquad \qquad\qquad \widehat{L}_1 = L_1 \, ,
  \cr
  \widehat{L}_2 &=& \frac{1}{2} [ (2 + \alpha \xi^{-1} ) L_2
   + \alpha\xi^{-1}  J  ]
    \, , \qquad\qquad \qquad \widehat{D} = D \, ,
\cr
\widehat{P}_0 &=& -\frac{1 }{\sqrt{2}}
 \Bigg\{-  \xi^{-1} \ln
 [ 1 +
 \xi \, P_{+}  +
 \frac{1}{2} \alpha \xi P_2 (1
 -
 \alpha \, P_{-})^{-1}]
+\ \alpha^{-1} \ln (1 +
\alpha \, P_{-} )\Bigg\}\, ,
    \cr
\widehat{P}_1 &= &\frac{ 1}{\sqrt{2}} \Bigg\{ \xi^{-1} \ln
 [ 1 +
 \xi \, P_{+} \,
 +
 \frac{1}{2} \alpha \xi P_2 (1
 -
 \alpha \, P_{-})^{-1}]
  + \ \alpha^{-1} \ln (1 +
 \alpha \, P_{-}
 )\Bigg\}\, ,
\cr
    \widehat{P}_2 &= & P_2
      [ 1 +
      \alpha \, P_{-} ]^{-1} \, ,
\cr
\widehat{K}_0 & = & K_0 +
 \frac{1 }{\sqrt{2}} \Bigg[ - \alpha DL_1 +
\frac{1 }{4} \xi (D- L_1)^2
 -
  \alpha
\, \xi^{-1}
 \, K_{-}
 -
\frac{\alpha }{4}(L_2 +J)^2\Bigg]\, ,
\cr
\widehat{K}_1 & = & K_1 + \frac{1 }{\sqrt{2}}
\Bigg[ - \alpha DL_1 -
\frac{1 }{4} \xi (D-L_1)^2
-
 \, \alpha
\, \xi^{-1}
 \, K_{-}
+
\frac{\alpha }{4}(L_2 +J)^2\Bigg]\, ,
\cr
\widehat{K}_2
 & = &  K_2 -
 \frac{\alpha}{\sqrt{2}} D(L_2 + J) \, ,
 \end{eqnarray}
Applying the formulae (3.10) to deformed generators we can
express the relations (2.15--17) in terms of deformed D=3
 conformal generators (\ref{lulymo3.11a})

ii) {\bf D=4 AdS basis}

Ten generators of $o(3,2)$ satisfying the
relations (\ref{llm22.9})  describe $D=4 \, AdS$ algebra defined by the
commutation relations

for D=4  Lorentz generators ($M_i , N_i; i=1,2,3 $)
\begin{equation}\label{llm3.12}
  [M_i , M_j ] = \epsilon_{ijk} \, M_k \, , \qquad
  [ N_i , N_j ] = -\epsilon_{ijk} \, M_k \, , \qquad
  [M_i , N_i ] = \epsilon_{ijk} \, N_k \, ,
\end{equation}

and for the extension by the four curved $AdS$ translations
${\cal P}_\mu = ({\cal P}_i , {\cal P}_0 )$
\begin{equation}\label{llm3.13}
\begin{array}{ll}
 [{\cal P}_i , {\cal P}_j ] = - \frac{1}{R^2} \, \epsilon_{ijk}\, M_k\, ,
& [{\cal P}_0 , {\cal P}_i ] =  \frac{1}{R^2} \, N_i\, ,
 \cr [M_i
, {\cal P}_0 ] = 0 \, , &[{M}_i , {\cal P}_j ] = \epsilon_{ijk}\,
{\cal P}_k\, ,
\cr [N_i ,{\cal  P}_0 ] =  {\cal P}_i \, , &[N_i , {\cal
P}_j] = \delta_{ij} \,   {\cal P}_0 \, ,
\end{array}
\end{equation}
where $R$ describes dimensionfull  (inverse mass dimension)
 $AdS$ radius.


  The D=4 $AdS$
generators ($M_i, N_i , {\cal P}_\mu $)
can be expressed in terms of "mathematical" generators $h_2, h_4, e_{\pm A}$
as follows

\begin{equation}\label{lulymo3.14}
\begin{array}{rlll}
M_1 &=&  \frac{1}{\sqrt{2}}(e_{-1} - e_1) \, , \qquad
&
 M_2 = \frac{1}{\sqrt{2}}(e_{-3} - e_3) \, ,
\cr
M_3 &=&   \frac{1}{{2}}(e_{-4} - e_2 + e_2 - e_4) \, ,
&
\cr
N_1 &=& - \frac{1}{{2}}(e_{-2} + e_4 + e_2 + e_4) \, ,
\qquad
&
N_2 =  \frac{1}{{2}}(h_{4} - h_2) \, ,
\cr
N_3 &=&  \frac{1}{\sqrt{2}}(e_{-1} + e_1) \, ,
&
\cr
{\cal P}_0 &=&  \frac{1}{{2R}}(e_{-2} + e_{-4} - e_2 - e_4) \, ,
\qquad
&
{\cal P}_1 = \frac{1}{{2R}}(h_{4} + h_{2} ) \, ,
\cr
{\cal P}_2 &=&  -\frac{1}{{2R}}(e_{-4} - e_{-2} - e_2 + e_4) \, ,
\qquad
&
{\cal P}_3 = \frac{1}{{2R}}(e_{-3} + e_{3} ) \, ,
\end{array}
\end{equation}

The inverse formulae to the relations (3.14) permit
 to express
 the classical $r$-matrix (\ref{llm2.6}) in D=4 $AdS$
   basis:

\begin{eqnarray}\label{lulymo3.16}
 r(\alpha, \xi, \rho; R)& = &\frac{1}{2} N_2 \wedge
  \Big[ (\xi -\alpha) N_1
 - (\alpha + \xi )M_3 \Big]
 +\alpha M_2 \wedge (N_3 - M_1 )
\cr
&&
  - \frac{1}{2}\rho M_3 \wedge N_1
 + \frac{R}{2}\Big\{ N_2 \wedge \Big[ (\xi -\alpha){\cal P}_0 -
(\alpha +\xi){\cal P}_2\Big]
\cr
&&
 + {\cal P}_1 \wedge \Big[(\xi-\alpha)M_3
 - (\alpha+\xi)N_1\Big]
- \rho
  M_3 \wedge {\cal P}_0
  \cr
  &&+ \, \rho\,  N_1 \wedge {\cal P}_2 - 2    \alpha {\cal P}_3
  \wedge (N_3 - M_1 )  \Big\}
\cr && + \, \frac{R^2}{2}\Big\{ {\cal P}_1 \wedge
\Big[(\xi-\alpha){\cal P}_2 -(\alpha+\xi){\cal P}_0\Big] - \rho
{\cal P}_0 \wedge {\cal P}_2 \Big\} \, .
\end{eqnarray}

We had already seen in the formulae (3.14) that in order to obtain the
physical rescaling of AdS generators
 ($P_{\mu} \longrightarrow \lambda^{-1}P_{\mu}$,
$M_i \longrightarrow M_i$, $N_i \longrightarrow N_i$) it is
sufficient to consider the $AdS$ radius $R$ as the dimensionfull
parameter ($R \longrightarrow \lambda R$),
 i.e. we need not
transform the generators $e_{+A},e_{-A}$, $h_r$ under the scaling.
Subsequently, from the formulae (2.6) as well as (3.15) it
follows that the deformation  parameters $\alpha, \xi, \rho$ in
D=4 $AdS$ basis (\ref{lulymo3.14}) remain dimensionless.

Using further the relations (3.14) one can express the twist
factor (2.11) as well as the twisted coproduct formulae from
 Appendix in terms of D=4 $AdS$ generators. The choice of
 deformation
  map introducing nonclassical basis
 suitably adjusted to D=4 $AdS$ algebra is under
 consideration.

\section{Outlook}
\setcounter{equation}{0}

\hspace{12pt} This paper contributes to the  studies of nonstandard quantum deformations of D=4
 space-time algebras which are described infinitesimally by the classical $r$-matrices
 satisfying CYBE. Our approach is based on Drinfeld twist technique, which provides
 the complete set of Hopf algebra relations and gives the universal
 ${\cal R}$-matrix via the formula
 ${\cal R}= {\cal F}^{21} {\cal F}^{-1}$. For example  performing  the
  twist quantization of the classical
   $r$-matrix (\ref{llm2.6}) with $\rho=0$
   we obtain
the universal
${\cal R}$-matrix in the deformed $o(3,2)$-basis
   which   looks as follows
   (see also
    \cite{ll16})
  \begin{eqnarray}\label{lulymo4.1}
{\cal R} &=&\exp \left( {E}_{2}\otimes H_{2}\right) \exp
\left( -\alpha E_{1}e^{-\frac{1}{2}{ E }_{4}}\otimes E_{2}\right)
\exp \left( { E }_{4}\otimes H_{4}\right)
\cr
&&\times \exp \left( -H_{4}\otimes {E }_{4}\right) \exp \left(
\alpha E_{2}\otimes E_{1}e^{-\frac{1}{2}{E}_{4}}\right) \exp
\left( -H_{2}\otimes {E }_{2}\right) .
\end{eqnarray}

Further, using the formulae (\ref{lulymo3.14}) and
    (\ref{llm2.17}) one can express the formula
(\ref{lulymo4.1}) in terms of D=3 conformal or D=4 $AdS$ generators.
Finally let us recall that
   the twist quantization of space-time symmetries in classical Lie algebra basis
   does not modify the irreducible representations of space-time algebra
   (one-particle sectors),
 but  provides deformed tensor   product representations, i.e. non-Fock
formula for $n$-particle states. Such a modification of tensoring
procedure of irreducible representations  can
    be interpreted as the introduction of particle interactions
    in algebraic way (see e.g.
     \cite{ll17,ll18}).
     Such interpretation of our  scheme follows  from
the fact that the Hamiltonian $H$ is the time component of momentum vector
 and belongs to the set of symmetry generators with deformed coproducts. If we
interpret  the coproduct $\Delta(H)$ as describing two-particle energy
operator, we obtain nonstandard formula for nonsymmetric two-particle
interaction energy which is not invariant under the
 classical   particle exchange transformation.
It is
 an interesting task
  to find the suitable definition of deformed statistics
 and its applications in physical models
  with \sloppy
   interaction terms appearing due to quantum deformations.

\subsection*{Acknowledgments}
The paper has been supported by KBN grant 5P03B05620 (JL+MM) and
the Russian Foundation for Fundamental Research under the grant N
00-01-00500. Also one of the authors (JL) wishes to acknowledge
the financial support
 of Valencia University and of grant BFM 2002-03681.

\section*{Appendix}

Substituting twist function (2.11) in the formula
(\ref{xxx}) we obtain the following coproducts for the generators
$h_2, h_4 = 2 h_1 + h_2, e_{\pm A}$ $(A=1,2,3,4)$ describing
$o(3,2)$ Lie algebra:

\begin{eqnarray*}
\begin{array}{ccc}
\Delta _{{\cal F}}\left( h_{4}\right)  & = &
\begin{array}{c}
h_{4}\otimes e^{-{\bf  \sigma }_{4} (\alpha) }+1\otimes h_{4}
\\
- 2 \alpha   e_1 \otimes  e_3 e^{-\frac{3}{2}{\bf  \sigma }_{4}
(\alpha) -\frac{1}{2}{\bf %
 \sigma }_{2} (\alpha, \xi ) }-\frac{\alpha \xi }{2} h_2 \otimes  e_3 ^{2}e^{-2{\bf %
 \sigma }_{4} (\alpha) -{\bf  \sigma }_{2} (\alpha, \xi ) },%
\end{array}
\\
\Delta _{{\cal F}}\left(  h_2 \right)  & = &  h_2 \otimes e^{-{\bf %
 \sigma }_{2} (\alpha, \xi ) }+1\otimes  h_2 ,
 \\
\Delta _{{\cal F}}\left(  e_4 \right)  & = &  e_4 \otimes e^{{\bf
 \sigma }_{4} (\alpha) }+1\otimes  e_4 ,
 \\
\Delta _{{\cal F}}\left(   e_1 \right)
& = &   e_1 \otimes e^{-\frac{1}{2}%
{\bf  \sigma }_{4} (\alpha) -\frac{1}{2}{\bf  \sigma }_{2}
(\alpha, \xi ) }+1\otimes
  e_1 - \frac{1}{2} \xi
 h_2 \otimes  e_3 e^{-{\bf  \sigma }_{4}
  (\alpha) -{\bf  \sigma }_{2} (\alpha, \xi ) },
 \\
\Delta _{{\cal F}}\left(  e_3 \right)
  & = &  e_3 \otimes e^{\frac{1}{2}{\bf %
 \sigma }_{4} (\alpha)
  +\frac{1}{2}{\bf  \sigma }_{2} (\alpha, \xi ) } +
 e^{{\bf \sigma }%
_{4}(\alpha)}\otimes  e_3 ,
\\
\Delta _{{\cal F}}\left(   e_2\right)  & = &
\begin{array}{c}
  e_2\otimes e^{{  \sigma }_{2} (\alpha, \xi ) }+1\otimes   e_2
  \\
+ \xi  e_3 \otimes  e_3 e^{-\frac{1}{2}
{  \sigma }_{4} (\alpha) +\frac{1}{2}{\bf %
 \sigma }_{2} (\alpha, \xi ) }
 + \frac{1}{2}\xi ^{2} e_4 \otimes  e_3 ^{2}e^{-{\bf \sigma }%
_{1+2}}.%
\end{array}%
\end{array}
\label{cop-pos}
\end{eqnarray*}%
\[
\begin{array}{l}
\Delta _{{\sf {\cal F}}}\left(   e_{-1} \right) =  e_{-1} \otimes e^{-\frac{1}{2}%
{\bf  \sigma }_{4} (\alpha) +\frac{1}{2}{\bf  \sigma }_{2}
(\alpha, \xi ) }+1\otimes   e_{-1}
\\
+ \alpha   e_1 \otimes \xi ^{-1}\left( e^{{\bf  \sigma }_{2}
 (\alpha, \xi ) }-1\right) e^{-%
\frac{1}{2}{\bf  \sigma }_{4} (\alpha) -\frac{1}{2}{\bf  \sigma
}_{2} (\alpha, \xi ) } - \frac{1}{2}\alpha
 h_2 \otimes  e_3 e^{-{\bf  \sigma }_{2}
 (\alpha, \xi ) }e^{-{\bf  \sigma }_{4} (\alpha) };%
\end{array}%
\]%
\[
\begin{array}{l}
\Delta _{{\cal F}}\left(  e_{-2} \right) = e_{-2} \otimes e^{-{\bf \sigma }%
_{2}(\sigma,\xi)}+1\otimes  e_{-2} + \frac{1}{2}\alpha  h_2
 e_1 \otimes  e_3\,  e^{-\frac{3}{2}%
\left( {\bf  \sigma }_{4} (\alpha) +{\bf  \sigma }_{2} (\alpha,
\xi ) \right) }
\\
 - \frac{1}{2} \xi  h_2 \otimes \left(
\begin{array}{c}
\left( 1-e^{-{\bf  \sigma }_{2} (\alpha, \xi ) }\right) -\frac{1}{2}\left( 1-e^{-{\bf %
 \sigma }_{4} (\alpha) }\right)  \\
+  h_2 +\frac{1}{2}\alpha \xi
e_3 ^{2}e^{-2{\bf  \sigma }_{4} (\alpha) }e^{-{\bf %
 \sigma }_{2} (\alpha, \xi ) }-
 \alpha  e_3   e_1 e^{-{\bf  \sigma }_{4} (\alpha) }%
\end{array}%
\right) e^{-{\bf  \sigma }_{2} (\alpha, \xi ) } \\
- \frac{1}{8} h_2 ^{2}\otimes \left(
\begin{array}{c}
2\left( e^{-{\bf  \sigma }_{2} (\alpha, \xi ) }-1\right)  \\
- \alpha \xi  e_3 ^{2}e^{-2{\bf  \sigma }_{4} (\alpha) }e^{-{\bf  \sigma }_{2} (\alpha, \xi ) }%
\end{array}%
\right) e^{-{\bf  \sigma }_{2} (\alpha, \xi ) }
\\
+ \alpha    e_1 \otimes   e_1 e^{-\frac{1}{2}\left( {\bf  \sigma }_{4} (\alpha) +{\bf %
 \sigma }_{2} (\alpha, \xi ) \right) }-\frac{1}{2}\alpha   e_1 ^{2}\otimes \left( 1-e^{-{\bf %
 \sigma }_{4} (\alpha) }\right) e^{-{\bf  \sigma }_{2} (\alpha, \xi ) };%
\end{array}%
\]
\[
\begin{array}{l}
\Delta _{{\cal F}}\left( e_{-3} \right)
 =e_{-3} \otimes e^{-\frac{1}{2}{\bf %
 \sigma }_{4} (\alpha) -\frac{1}{2}{\bf  \sigma }_{2}
  (\alpha, \xi ) }+1\otimes e_{-3}
 \\
+ \frac{1}{2}\xi  h_2 \otimes
\left(  - e_{-1} +
 \alpha  h _{3} e_3 e^{-{\bf \sigma }%
_{4}}-\frac{1}{2}\alpha ^{2} e_3 ^{2}  e_1 e^{-2{\bf  \sigma
}_{4} (\alpha) }\right) e^{-{\bf  \sigma }_{2} (\alpha, \xi ) }
\\
+ \frac{1}{2}\alpha \xi  h_2 \otimes  e_3
\left( -e^{-{\bf  \sigma }_{2} (\alpha, \xi ) }+\frac{1}{2}%
\alpha \xi  e_3 ^{2}e^{-2{\bf  \sigma }_{4}
(\alpha) -{\bf  \sigma }_{2} (\alpha, \xi ) }+\frac{1}{2}%
\right) e^{-\left( {\bf  \sigma }_{2} (\alpha, \xi ) +{\bf
\sigma }_{4} (\alpha) \right) }+
\\
 \frac{1}{4}\alpha \xi  h_2 ^{2}{\bf \otimes } e_3
  \left( e^{-{\bf  \sigma }_{2} (\alpha, \xi ) }-%
\frac{1}{2}\alpha \xi  e_3 ^{2}
e^{-2{\bf  \sigma }_{4} (\alpha)
 -{\bf \sigma }_{2}(\alpha,\xi)}-1\right) e^{-\left( {\bf  \sigma }_{2} (\alpha,
\xi ) +{\bf \sigma }_{4} (\alpha) \right) }
\\
- \frac{1}{2}\alpha  h_2   e_1 \otimes \left( -e^{-{\bf  \sigma }_{2}
(\alpha, \xi ) }+\frac{3}{2}%
\alpha \xi  e_3 ^{2}e^{-2{\bf  \sigma }_{4} (\alpha) }e^{-{\bf \sigma }%
_{2}(\alpha,\xi)}+1\right) e^{-\frac{1}{2}\left( {\bf  \sigma }_{2}
(\alpha, \xi ) +{\bf \sigma }%
_{4}(\alpha)\right) }
\\
+ \frac{1}{2}\alpha  h _{4}\otimes   e_1 e^{-{\bf  \sigma }_{4}
(\alpha) }+ \frac{1}{4}\alpha \xi
 h _{4} h_2 \otimes  e_3 e^{-2{\bf  \sigma }_{4} (\alpha) }
 \,
 e^{-{\bf \sigma }_{2}{(\alpha,\xi)}}
 -
\alpha  e_{-2}
 \otimes
  e_3\,
  e^{-{\bf  \sigma}_{4}(\alpha)}
 \, e^{-{\bf \sigma }_{2}(\alpha, \xi)}
\\
+
\alpha   e_1 \otimes  h _{3}e^{-\frac{1}{2}
\left( {\bf  \sigma }_{2} (\alpha, \xi ) +{\bf %
 \sigma }_{4} (\alpha) \right) }
 - \frac{1}{2}
 \alpha  h _{4} e_1 \otimes \left( e^{-\frac{1}{2}%
{\bf  \sigma }_{4} (\alpha) }-e^{-\frac{3}{2}{\bf  \sigma }_{4} (\alpha) }\right) e^{-\frac{1}{%
2}{\bf  \sigma }_{2} (\alpha, \xi ) }
\\
- \alpha ^{2}  e_1 \otimes  e_3  e_1\,
 e^{-\frac{3}{2}{\bf  \sigma }_{4} (\alpha) -\frac{1%
}{2}{\bf  \sigma }_{2} (\alpha, \xi ) }+\alpha ^{2}  e_1 ^{2}\otimes  e_3 \left( \frac{1}{2}%
-e^{-{\bf  \sigma }_{4} (\alpha) }\right) e^{-{\bf  \sigma }_{4} (\alpha) }e^{-{\bf \sigma }%
_{2}(\alpha,\xi)};%
\end{array}%
\]%
\[
\begin{array}{l}
\Delta {\sf _{{\cal F}}}\left(  e_{-4}\right) = e_{-4}\otimes e^{-{\bf %
 \sigma }_{4} (\alpha) }+1\otimes  e_{-4}
 \\
\frac{1}{2}\,  h_2 \otimes \left[
\begin{array}{c}
 + \left( \alpha \xi  e_3   e_{-1}
 +\frac{1}{2}\alpha \xi   e_2\right) e^{-{\bf %
 \sigma }_{4} (\alpha) -{\bf  \sigma }_{2} (\alpha, \xi ) }
 \\
-\frac{1}{2}\alpha ^{2}\xi  e_3 ^{2}\left( e^{{\bf  \sigma }_{2}
(\alpha, \xi ) }-1\right) e^{-2{\bf  \sigma }_{4} (\alpha)
-2{\bf  \sigma }_{2} (\alpha, \xi ) }
\\
+\alpha ^{2}\xi  e_3 ^{2}\left( e^{-3{\bf  \sigma }_{4} (\alpha) -{\bf \sigma }%
_{2}(\alpha,\xi)}-\frac{1}{4}\alpha \xi  e_3 ^{2}e^{-4{\bf  \sigma
}_{4} (\alpha) -2{\bf  \sigma }_{2} (\alpha, \xi ) }\right)
\end{array}%
\right] +
\\
+\frac{1}{8}\alpha ^{2}\xi  h_2 ^{2}\otimes  e_3 ^{2}\left(
e^{{\bf  \sigma }_{2} (\alpha, \xi ) }-1+\frac{1}{2}\alpha \xi
 e_3 ^{2}e^{-2{\bf  \sigma }_{4} (\alpha) }\right)
e^{-2{\bf  \sigma }_{4} (\alpha) }e^{-2{\bf  \sigma }_{2}
(\alpha, \xi ) }+
\\
- \alpha e_{-3} \otimes  e_3 e^{-\frac{3}{2}{\bf  \sigma }_{4} (\alpha) }e^{-\frac{1}{2}%
{\bf  \sigma }_{2} (\alpha, \xi ) }+\alpha   e_1 \otimes
e_{-1} e^{-\frac{1}{2}{\bf \sigma }%
_{1+2}}e^{-\frac{1}{2}{\bf  \sigma }_{2} (\alpha, \xi ) }
\\
+ \frac{1}{2}\alpha ^{2} h_2   e_1 \otimes  e_3 \left( e^{{\bf
\sigma }_{2} (\alpha, \xi ) }-1+\alpha
\xi  e_3 ^{2}e^{-2{\bf  \sigma }_{4} (\alpha) }\right) e^{-\frac{3}{2}{\bf \sigma }%
_{1+2}}e^{-\frac{3}{2}{\bf  \sigma }_{2} (\alpha, \xi ) }
\\
+\alpha ^{2}  e_1 ^{2}\otimes
\left( -\frac{1}{2} e_2
\,
 e^{-{\bf \sigma }%
_{1+2}}e^{-{\bf  \sigma }_{2} (\alpha, \xi ) }+
\alpha  e_3 ^{2}e^{-3{\bf  \sigma }_{4} (\alpha) -%
{\bf  \sigma }_{2} (\alpha, \xi ) }\right)
\\
+ \frac{1}{2}\alpha ^{2} e_{-2} \otimes  e_3 ^{2}e^{-2{\bf  \sigma }_{4} (\alpha) }e^{-%
{\bf  \sigma }_{2} (\alpha, \xi ) }
\\
-\alpha^2 \left( \frac{1}{2} h_4 \, e_1 + e_1\right)
 \otimes  e_3 \left( 2e^{-{\bf %
 \sigma }_{4} (\alpha) }-1\right) e^{-\frac{3}{2}{\bf  \sigma }_{4}
 (\alpha) }e^{-\frac{1}{2}%
{\bf  \sigma }_{2} (\alpha, \xi ) }
\\
+\frac{1}{2} h _{4}\otimes  h _{4}e^{-{\bf  \sigma }_{4} (\alpha)
}+
 \frac{1}{2} h _{4}\otimes \left( e^{-%
{\bf  \sigma }_{4} (\alpha) }-e^{-2{\bf  \sigma }_{4} (\alpha)
}\right)
\\
- \alpha ^{2}  e_1 \otimes  h _{4} e_3 e^{-\frac{3}{2}{\bf  \sigma }_{4} (\alpha) -%
\frac{1}{2}{\bf  \sigma }_{2} (\alpha, \xi ) }
\\
 - \frac{1}{4}\alpha ^{2}\xi(  h _{4} h_2
 \otimes  e_3 ^{2}e^{-3{\bf  \sigma }_{4}
(\alpha) -{\bf %
 \sigma }_{2} (\alpha, \xi ) }
 +  h_2 \otimes  h _{4} e_3 ^{2}e^{-2{\bf %
+ \sigma }_{4} (\alpha) -{\bf  \sigma }_{2} (\alpha, \xi ) }).%
\end{array}%
\]%
where $h_3 = h_1 + h_2 = \frac{1}{2}(h_2 + h_4 )$.


\begin{thebibliography}{99}

\bibitem{ll1} J. Lukierski, P. Minnaert and M. Mozrzymas, Phys.
 Lett. {\bf B371}, 215 (1996).


 \bibitem{ll2} V.G. Drinfeld, "Quantum Groups", in Proc. of XX-th
 Inern. Congress of Math. (Berkeley, USA, 1986) p. 798.



 \bibitem{ll3} M. Jimbo, Lett. Math. Phys. {\bf 10}, 63
  (1985).



\bibitem{ll5} V.K. Dobrev, J. Phys. {\bf A26}, 1317 (1993).



\bibitem{ll4} J. Lukierski, A. Nowicki, Phys. Lett. {\bf
B279}, 299 (1992).





 \bibitem{ll6} A. Ballesteros, F.J. Herranz, M.A. del Olmo and M.
 Santander, J. Phys. {\bf A28}, 941 (1995).

\bibitem{ll7} J. Lukierski, V. Lyakhovsky and M. Mozrzymas, Phys.
 Lett. {\bf B538}, 375 (2002).


\bibitem{ll8} F.J. Herranz, J. Phys. {\bf A30}, 6123 (1997).

\bibitem{ll9} V.G. Drinfeld, Dokl. Acad. Nauk SSSR {\bf 273}, 531
(1983).


\bibitem{ll10} P.P. Kulish, V.D. Lyakhovsky, A.I. Mudrov,
J. Math. Phys. {\bf 40}, 4569 (1999).


\bibitem{ll11} P. P. Kulish, V. D. Lyakhovsky
 and M. A. del Olmo, J. Phys. A: Math. Gen. {\bf 32}, 8671 (1999).

 \bibitem{ll12} J. Lukierski, A. Nowicki, H. Ruegg,
 Phys. Lett. {\bf B271}, 321 (1991).




\bibitem{ll13} S. Majid and H. Ruegg,
 Phys. Lett.  {\bf B334}, 348 (1994).

\bibitem{ll14} J. Lukierski, H. Ruegg, W.J. Zakrzewski, Ann.
 Phys. {\bf 243}, 90 (1995).




\bibitem{ll15} P. P. Kulish and V. D. Lyakhovsky, J. Phys. A: Math. Gen. {\bf %
33}, L279 (2000).

\bibitem{ll16} P. P. Kulish, V. D. Lyakhovsky and A. A. Stolin, J. Math.
Phys. {\bf 42}, 5006 (2001).

\bibitem{ll17} J. Lukierski, A. Nowicki, H. Ruegg, V.N. Tolstoy,
J. Phys. {\bf A27}, 2383 (1994).

\bibitem{ll18} J. Lukierski, A. Nowicki, in Proceedings of Quantum Group
 Symposium at "Group21", Eds. H.-D. Doebner and V.K. Dobrev, Heron Press,
 Sofia, 1997, p. 173.

\end{thebibliography}
\end{document}